\documentclass[conference]{IEEEtran}
\IEEEoverridecommandlockouts

\usepackage{amsmath,amssymb,amsfonts}
\usepackage{algorithmic}
\usepackage{graphicx}
\usepackage{textcomp}
\usepackage{xcolor}
\usepackage[a4paper, total={184mm,239mm}]{geometry}
\usepackage{mathptmx} 
\usepackage{fancyhdr}
\usepackage[normalem]{ulem}
\usepackage[hyphens]{url}
\usepackage[sort,nocompress]{cite}
\usepackage[final]{microtype}
\usepackage[keeplastbox]{flushend}
\usepackage[linesnumbered,ruled]{algorithm2e}
\usepackage{amsmath}
\usepackage{booktabs}
\usepackage{multirow}
\usepackage{array}
\usepackage{tabularx}
\usepackage{color}
\usepackage{enumerate, pifont}
\usepackage{diagbox}
\usepackage{enumitem}
\usepackage{enumerate, pifont}

\def\BibTeX{{\rm B\kern-.05em{\sc i\kern-.025em b}\kern-.08em
    T\kern-.1667em\lower.7ex\hbox{E}\kern-.125emX}}

\begin{document}

\title{PiPoMonitor: Mitigating Cross-core Cache Attacks Using the Auto-Cuckoo Filter}

\author{\IEEEauthorblockN{Fengkai Yuan\IEEEauthorrefmark{1},Kai Wang\IEEEauthorrefmark{2}, Rui Hou\IEEEauthorrefmark{1}\IEEEauthorrefmark{3}\thanks{\IEEEauthorrefmark{3}Corresponding Author: Rui Hou}, Xiaoxin Li\IEEEauthorrefmark{1}, Peinan Li\IEEEauthorrefmark{1}, Lutan Zhao\IEEEauthorrefmark{1}, Jiameng Ying\IEEEauthorrefmark{1}, Amro Awad\IEEEauthorrefmark{4}, Dan Meng\IEEEauthorrefmark{1}}
\IEEEauthorblockA{\IEEEauthorrefmark{1}\textit{State Key Laboratory of Information Security, Institute of Information Engineering, Beijing, China}}
\IEEEauthorblockA{\IEEEauthorrefmark{2}\textit{Department of Computer Science and Technology, Harbin Institute of Technology, Harbin, China}}
\IEEEauthorblockA{\IEEEauthorrefmark{4}\textit{North Carolina State University, North Carolina, USA}}
\IEEEauthorblockA{\IEEEauthorrefmark{1}\textit{\{yuanfengkai, hourui, lixiaoxin, lipeinan, zhaolutan, yingjiameng\}@iie.ac.cn}}
\IEEEauthorblockA{\IEEEauthorrefmark{2}\textit{wk1220ym@163.com}, \IEEEauthorrefmark{4}\textit{ajawad@ncsu.edu}}
}

\maketitle

\begin{abstract}

Cache side channel attacks obtain victim cache line access footprint to infer security-critical information. Among them, cross-core attacks exploiting the shared last level cache are more threatening as their simplicity to set up and high capacity.
Stateful approaches of detection-based mitigation observe precise cache behaviors and protect specific cache lines that are suspected of being attacked. However, their recording structures incur large storage overhead and are vulnerable to reverse engineering attacks. Exploring the intrinsic non-determinate layout of a traditional Cuckoo filter, this paper proposes a space efficient Auto-Cuckoo filter to record access footprints, which succeed to decrease storage overhead and resist reverse engineering attacks at the same time. With Auto-Cuckoo filter, we propose PiPoMonitor to detect \textit{Ping-Pong patterns} and prefetch specific cache line to interfere with adversaries' cache probes.
Security analysis shows the PiPoMonitor can effectively mitigate cross-core attacks and the Auto-Cuckoo filter is immune to reverse engineering attacks.
Evaluation results indicate PiPoMonitor has negligible impact on performance and the storage overhead is only 0.37$\%$, an order of magnitude lower than previous stateful approaches.
\end{abstract}


\section{Introduction}


Cache hierarchy is one of the most critical designs to improve performance in commercial processors.
However, the execution footprints left on different cache levels can be observed by attackers to infer security-critical information.
In particular, cross-core attacks on the last level cache (LLC) are dangerous and has been widely exploited due to its high bandwidth, low noise and low construction difficulty\cite{DBLP:conf/dac/KayaalpAPJ16}.

There are three orthogonal kinds of mitigation against cross-core attacks:
\textit{\ding{182}Partition-based mechanisms} physically isolate cache resources to protect victim process from attackers in different domains;
\textit{\ding{183}Randomization-based mechanisms} change the cache layout periodically to confuse attackers to infer secrets, which mitigate attacks in both intra- and inter- security domain;
\textit{\ding{184}Detection-based mechanisms} monitor the cache behaviors to detect suspicious operations, providing the information when there may exist attacks.
The first two mechanisms have been well studied.
This paper focuses on the third mechanism.

Existing detection-based defenses fall into three categories. 
\ding{172} Machine Learning-based approaches identify malicious operations based on machine learning algorithms.
The accuracy of this method largely depends on the test suites and it cannot figure out the root cause of the information leakage.
\ding{173} Stateless approaches\cite{DBLP:conf/dac/KayaalpKEEAPJ17,DBLP:conf/isca/YanGST17,DBLP:conf/IEEEpact/Panda19} observe the events of back-invalidations in private cache lines from LLC evictions. 
However, back-invalidations vastly exist in benign execution, resulting in high false positives and performance overhead.
\ding{174} Stateful approaches\cite{DBLP:conf/date/WangYHJM20, DBLP:conf/cf/WangYHLJM19} avoid high false positives by observing more precise cache behaviors. They introduce extra structure to record suspicious temporal correlated LLC-memory interactions (called \textit{Ping-Pong patterns}) of the same cache line.
However, recording long history to capture the suspicious access pattern results in large storage overhead.
Even worse, the limited-size recording structure is vulnerable to reverse engineering attacks.

In this paper, we propose a novel stateful approach with low false positives, low storage overhead and resistance to reverse engineering attacks, named as PiPoMonitor.
It captures \textit{Ping-Pong pattern} by monitoring the traffic history between LLC and memory.
Once detecting abnormal behaviors, the corresponding cache lines are prefetched to interfere the secrets inference.
To minimize the storage overhead and resist reverse engineering attacks, we deploy cuckoo filter to record traffic since it record the short hash results of addresses instead of themselves.
However, attackers can delete certain records in traditional Cuckoo filter to bypass detection, resulting in false negatives.
Thus, we propose the Auto-Cuckoo filter, which autonomically deletes the record stored in the i-th relocated destination where i is a predefined threshold.
The autonomic deletion exponentially increases the uncertainty of a record eviction, rendering reverse engineering attacks impractical.

This paper makes the following contributions:

\begin{enumerate}
    \item We first propose PiPoMonitor to record traffic history between LLC and memory with space efficient hardware Cuckoo filter to detect \textit{Ping-Pong patters} in cross-core cache side-channel attacks effectively.

    \item Based on traditional Cuckoo filter, we implement an Auto-Cuckoo filter with autonomic deletion to prevent malicious bypass in PiPoMonitor aware attacks.

    \item After detecting malicious behaviors, PiPoMonitor prefetches specific cache lines to confuse the attackers.


\end{enumerate}

\section{Background}
\label{Background}
\subsection{Ping-Pong Patterns}


To infer secrets of victim through cross-core cache side-channel attacks, attackers iteratively probe the target secret-dependent cache lines, causing frequent migration back and forth between LLC and memory.
These cache line access behaviors under attack are defined as \textit{Ping-Pong patterns}\cite{DBLP:conf/date/WangYHJM20, DBLP:conf/cf/WangYHLJM19}. 

\begin{figure}[h] 
	\centering
	\includegraphics[width=0.32\textwidth]{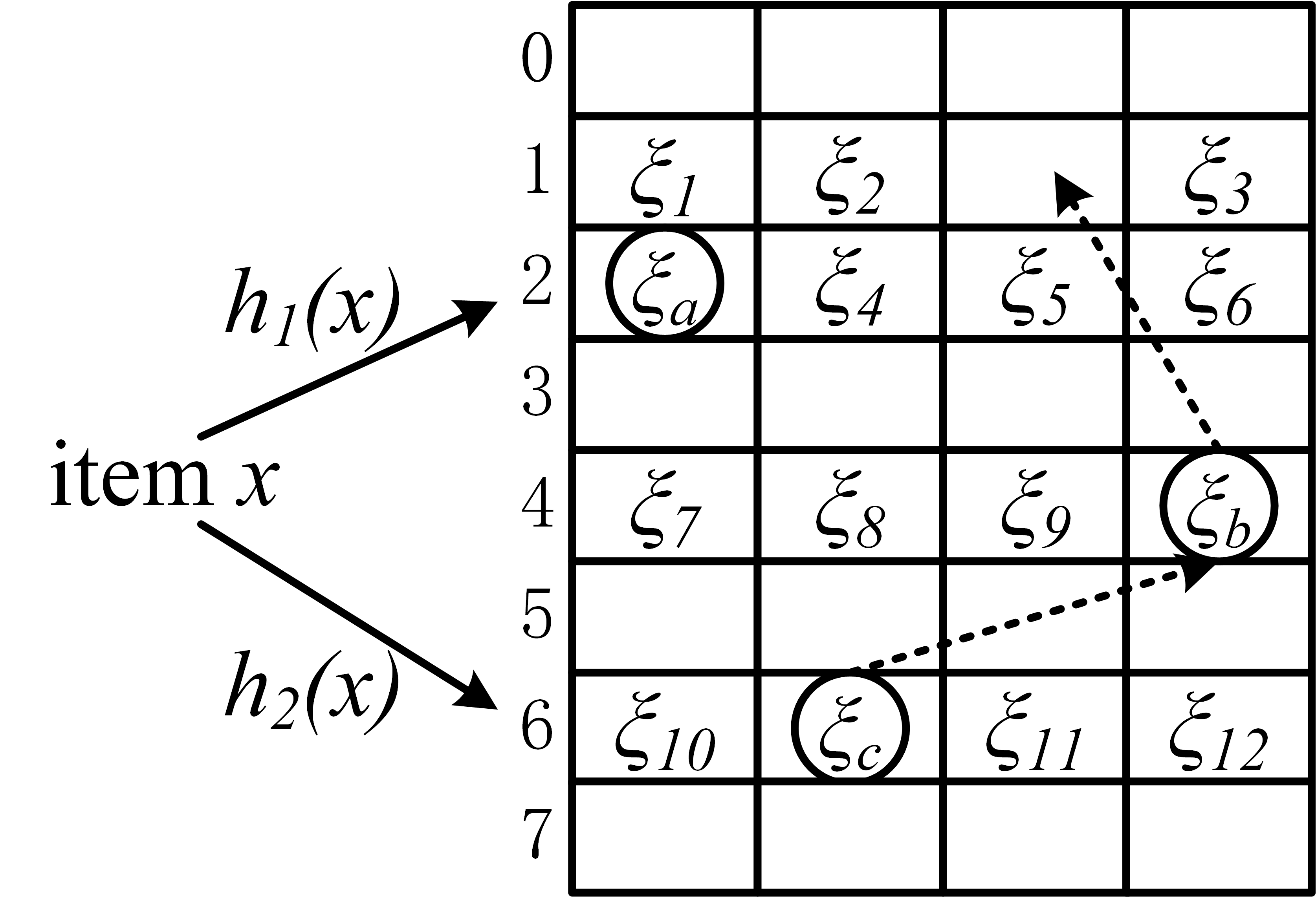}
	\caption{Record relocations of the Cuckoo filter.}
	\label{fig:cuckooFilterBasics}
	\vspace{-1.0em}
\end{figure}

\subsection{Cuckoo Filter}

A traditional Cuckoo filter \cite{DBLP:conf/conext/FanAKM14} is a software data structure with three significant features:

\textbf{Space-efficiency:} It stores compact fingerprint records instead of the raw items.
As shown in Fig~\ref{fig:cuckooFilterBasics}, the filter first computes two candidate buckets indices for each item \textit{x} with following two hashing functions.
Then, it records the hash results as the fingerprints, dubbed $\xi$\textit{x}, to a matrix.

\textbf{Probabilistic false positives:}
Since hash results of different items may be the same, a Cuckoo filter falsely responds that a match exists with a bounded possibility for its specifications.

\textbf{Non-determinate Layout:} 
For a high occupancy, Cuckoo filters relocate records to search for more vacancies.
While inserting a new item, a filter allocates a entry in one of the two candidate bucket rows.
If all entries in the two bucket rows have already been occupied, the filter then selects a record randomly to relocate it to its alternative bucket row.
As Cuckoo filters only store fingerprints, it uses the partial-key Cuckoo hashing to compute bucket indices:
$$
\begin{aligned}
&h_1(x) = hash(x)\\
&h_2(x) = h_1(x) \oplus hash(\xi\textit{x})
\end{aligned}
$$
Thus, the alternative bucket index of a relocated record can be computed by performing an XOR based on current bucket and the fingerprint.
If the corresponding bucket rows are full, it has to select one more victim record and relocate it. 
Shown by Fig. \ref{fig:cuckooFilterBasics}, bucket row 6 is full while inserting item \textit{x}. Then record $\xi$\textit{c} is selected to relocate to its alternative candidate bucket row 4.
However, bucket row 4 is full as well, randomly selected $\xi$\textit{b} is relocated to bucket row 1 and find a vacancy.
Cuckoo filters specify an upper bound of the number of relocations (maximal number of kicks, denoted as MNK) to indicate that the filter is full.
When the filter is full, Cuckoo filters support deletion operations to remove some records manually.

\begin{table}[h]
\centering
\caption{Notations of the Auto-Cuckoo filter}
\begin{tabular}{cl} 
\toprule
Notations & Description \\
\midrule
\textit{x} & a cache line address\\
$\xi$\textit{x} & fingerprint of \textit{x} \\
$\mu$\textit{x}, $\sigma$\textit{x} & two candidate buckets of \textit{x} \\
\textit{l} & number of buckets \\
\textit{b} & number of entries per bucket \\
\textit{f} & length of fingerprint in bits \\
\textit{$\epsilon$} & false positive rate \\
\textit{MNK} & the maximal number of kicks \\
\textit{fPrint} & fingerprint field of a filter entry \\
\textit{Security} & reAccess counter of a filter entry \\
\textit{secThr} & the saturation value of \textit{Security} \\
\bottomrule
\end{tabular}
\label{tbl:table1}
\end{table}
\section{Threat Model}


PiPoMonitor aims at mitigating cross-core last-level cache side channel attacks.
We assume that adversaries are not privileged and do not share a physical core with the victim.
Our current implementation focus on attacks on inclusive LLC.
With PiPoMonitor guarding LLC-memory traffic, non-inclusive LLC can be slightly extended to defend against directory attacks \cite{DBLP:conf/sp/YanSGFCT19}.
\begin{figure}[h] 
	\centering
	\includegraphics[width=0.45\textwidth]{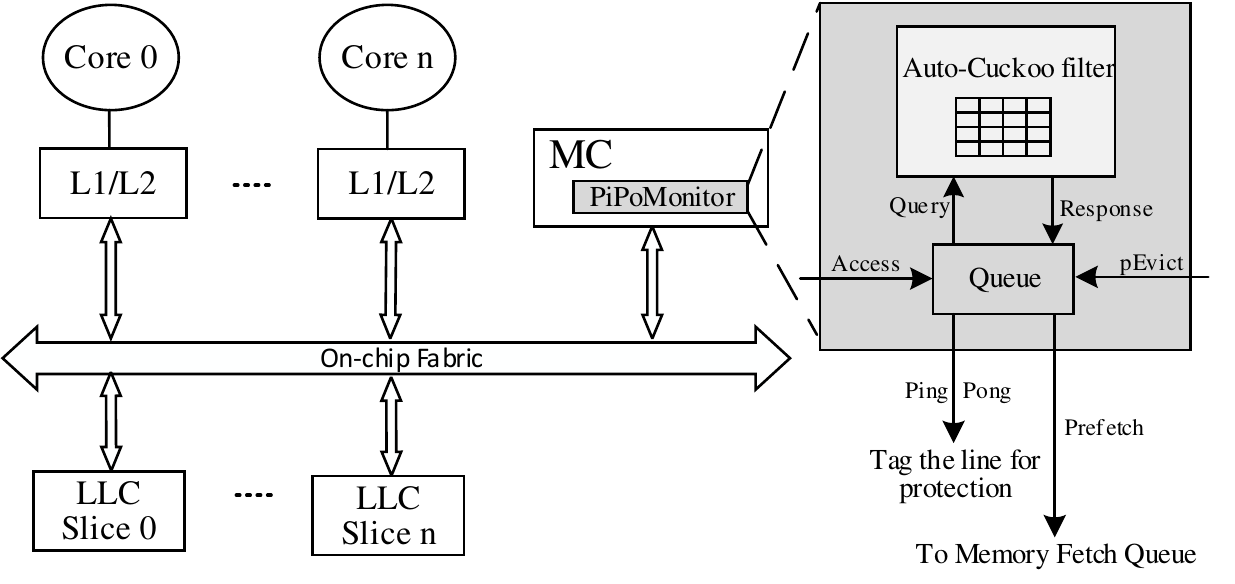}
	\caption{Architecture of the proposed approach}
	\label{fig:systemOverview}
	\vspace{-1.5em}
\end{figure}

\section{Overview of PiPoMonitor}
\label{sec_overview}
As shown in Figure~\ref{fig:systemOverview}, PiPoMonitor locates inside the on-chip memory controller (MC) and observes the memory access requests from LLC without extra network traffics.
Deployed with a hardware Auto-Cuckoo filter based on the Cuckoo filter, PiPoMonitor records cache line re-accesses and identifies lines behaving in \textit{Ping-Pong patterns} (Ping-Pong Lines).
Note that the PiPoMonitor works in parallel with memory fetches. 
It helps to hide latency of Response-waiting and avoid a critical path.
To demonstrate the Auto-Cuckoo filter conveniently, Table \ref{tbl:table1} summaries the notations and their descriptions. 

\textbf{Capturing Ping-Pong lines.}
The Auto-Cuckoo filter records the fingerprints of \textit{access requests to main memory} (\textbf{\textit{Access}}) in the buckets.
We use \textbf{\textit{reAccess}} to define the Access for the recorded line.
For each recorded line, the field of \textit{Security} to count the Access for the same line.
For an Access \textit{x} Query, the Auto-Cuckoo filter first checks whether there is a valid entry of $\xi\textit{x}$ in the buckets $\mu\textit{x}$ or $\sigma\textit{x}$.
If there is no valid entry of \textit{x}, the Auto-Cuckoo filter inserts a new entry in one of its candidate buckets, and its \textit{Security} is initialized as zero and returned to Response.
If the result is `exist', the \textit{Security} of the entry is increased by 1 to count reAccess and it is returned to Response.
When \textit{Security} counts up to the pre-defined \textit{secThr}, it satisfies the Ping-Pong pattern and \textit{x} will be captured as a Ping-Pong.




\textbf{Prefetching Ping-Pong lines.}
Once a Ping-Pong line is captured, the cache line will be tagged as Ping-Pong in LLC when it is retrieved to MC from memory.
When a tagged line is evicted, LLC will send a \textbf{\textit{pEvict}} message to PiPoMonitor.
After receiving the message, the latter waits for a pre-defined delay, and then sends a request to the memory fetch queue of MC to prefetch the line back to LLC, obfuscating adversaries' cache probes.
The delay is to avoid memory bandwidth preemption with the writeback of the same line.

Noting that once a Ping-Pong line has undergone Prefetch, it is tagged and will record whether it has been accessed. Only when the tagged-accessed line is evicted, it will be prefetched, so as to avoid over-protection by endless prefetching.







\linespread{1.0}

\section{Implementation}
\label{TheCuckooFilter}

\subsection{Autonomic Deletion}

The deletion of a classic Cuckoo filter is vulnerable. Because of false positives, addresses controlled by an adversary may have the same fingerprint and candidate buckets as the target line. Adversaries can exploit false deletions to remove one fingerprint (record) by deleting another under control.

We introduce autonomic deletions to modify insertions by making them never fails. When the number of relocations reaches \textit{MNK}, the Auto-Cuckoo filter evicts the last fingerprint that needs to be relocated. Recall that the fingerprint kicked out is randomly selected, and the alternative bucket for each fingerprint is different. The random kick paths inevitably lead to rich randomness of the fingerprint that is eventually evicted, which essentially increases the difficulty of constructing an eviction set by reverse attacks (proof in Section \ref{SecurityAnalysis}).

\begin{figure}[h] 
    \centering
	\includegraphics[width=0.46\textwidth]{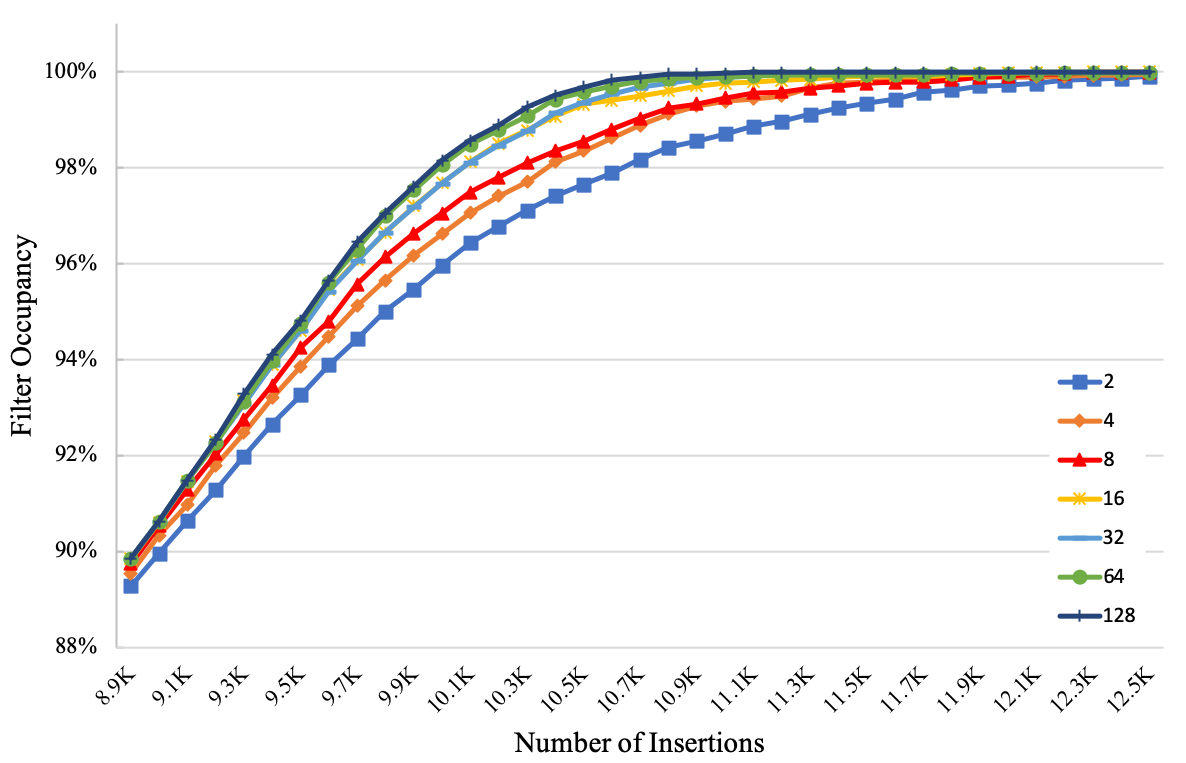}
	\caption{The occupancy of the Auto-Cuckoo filter using different \textit{MNK}}
	\label{fig:filterOccupancy}
	\vspace{-0.8em}
\end{figure}

Besides, autonomic deletion substantially decreases the \textit{MNK} to reach a high occupancy. \textit{MNK} is set as 100$\sim$1000 to avoid insertion failures in classic software Cuckoo filters.
For hardware implementation, a large \textit{MNK} means excessive hash computation and prohibitive relocation overhead.
In response, the Auto-Cuckoo filter relies on historical insertions to continuously find vacancies such that the occupancy climbs to 100$\%$, rather than counting on enormous relocations for each insertion.
Figure \ref{fig:filterOccupancy} depicts the occupancy as the insertion number increases. The filter parameters follow the configuration in Table \ref{tab:Setup}. We randomly pick addresses from memory address space and insert them into the filter using different \textit{MNK}. The results show the occupancy is not sensitive to \textit{MNK}. When the insertion number is less than 9K, the occupancy is even identical. Even if \textit{MNK} = 2, the occupancy can reach 100$\%$ when insertion number is 12.5K. We choose \textit{MNK} = 4 considering the trade-off between performance and security (see Section \ref{SecurityAnalysis}).

\begin{figure}[h] 
    \centering
	\includegraphics[width=0.46\textwidth]{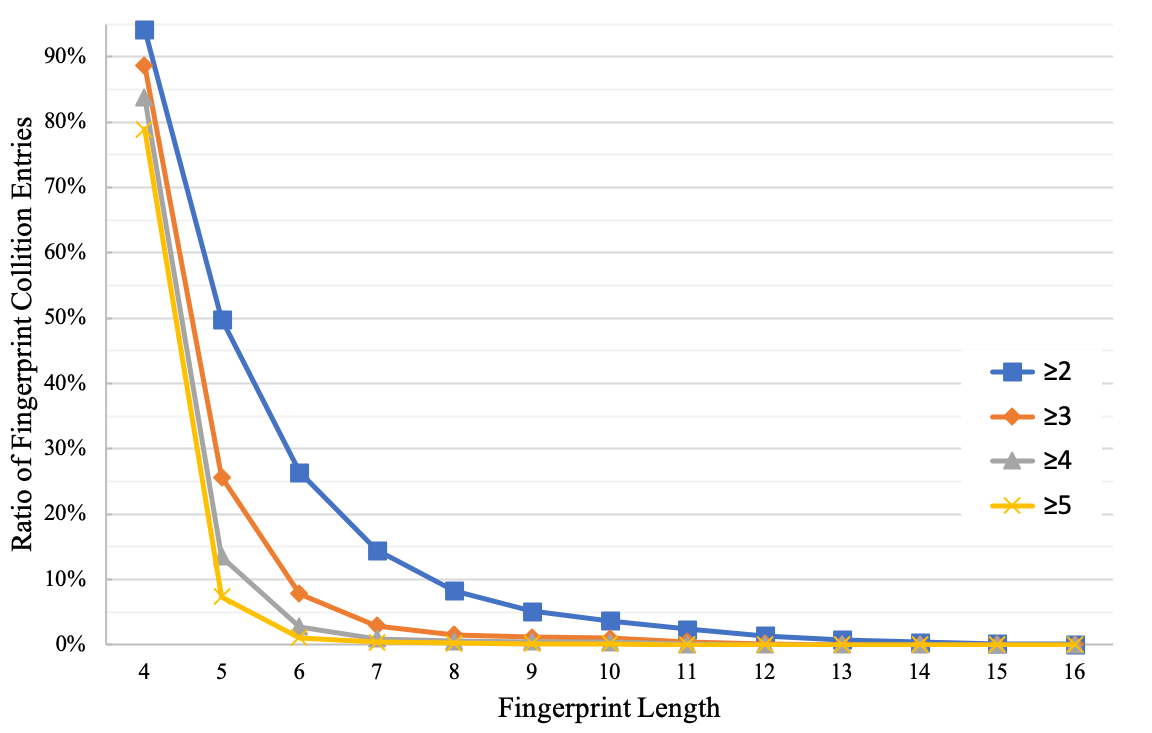}
	\caption{The ratio of fingerprint collision entries in the \textit{b}=8 Auto-Cuckoo filter with different \textit{f}. The result is classified according to the number of addresses that have collided in the entries}
	\label{fig:filterFalsePositiveRate}
	\vspace{-1.2em}
\end{figure}

\subsection{False Positive Rate}
The false positive rate, \textit{$\epsilon$}, of the Auto-Cuckoo filter directly affects the detection precision. A false positive occurs when different addresses having the same  fingerprint collide in their candidate buckets. 
 When a fingerprint collision occurs during a Query, records are merged into one entry with \textit{Security} increased. This accelerate the increase of \textit{Security} to \textit{secThr}, so that collision addresses are incorrectly captured as Ping-Pong.

Fortunately, the Auto-Cuckoo filter is superior to maintain a low \textit{$\epsilon$} with a small storage overhead. When looking up a non-existent record, a query checks two candidate buckets each with \textit{b} entries. For each entry, the probability that $\xi$\textit{x} is matched with \textit{fPrint} is 1/2${^{f}}$. After \textit{2b} comparisons, the upper bound of the total probability of a fingerprint collision is\cite{DBLP:conf/conext/FanAKM14}:
$$\epsilon = 1 - (1 - 1/2^{f})^{2b} \approx 2b/2^f$$

With \textit{l} and \textit{b} determined through security analysis in Section \ref{SecurityAnalysis}, the storage overhead of Auto-Cuckoo filter is \textbf{linear} with \textit{f}. In contrast, above equation shows that \textit{$\epsilon$} decreases \textbf{exponentially} as \textit{f} increases.
This means the filter can maintain a low false positive rate with a small storage overhead.
Fig. \ref{fig:filterFalsePositiveRate} shows after 6 million insertions, the ratios of entries with fingerprint collisions in the filter as \textit{f} increases. The ratio changes with \textit{f} approximately conforms to the equation. We choose \textit{f}=12 making the ratio as low as 0.014 with $\epsilon=0.004$. Moreover, shown in Fig. \ref{fig:filterFalsePositiveRate}, the ratio of the entries with more than 2 collision addresses approach 0 in this configuration.

\begin{figure}[h] 
    \centering
	\includegraphics[width=0.45\textwidth]{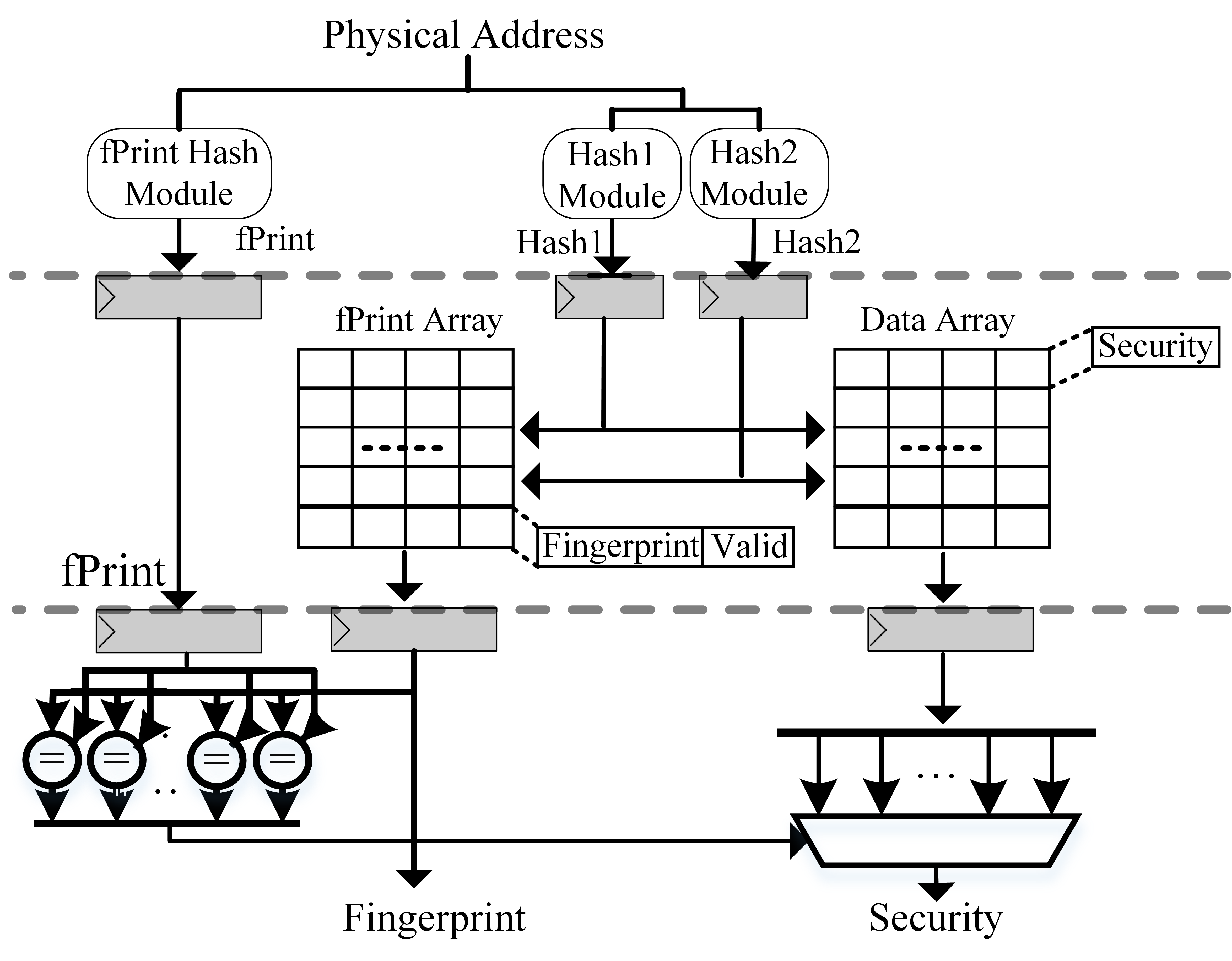}
	\caption{Hardware Microarchitecture of the Auto-Cuckoo filter}
	\label{fig:microArchitecure}
	\vspace{-2em}
\end{figure}

\subsection{Hardware Microarchitecture}
The microarchitecture of the Auto-Cuckoo filter is shown in Figure \ref{fig:microArchitecure}.
Similar to cache, the filter is composed of \textit{fPrint Array} and \textit{data Array}.
Correspondingly, each array has \textit{l} sets and each set has \textit{b} entries.
Differently, each physical address has two candidate buckets to be recorded. Hash1 Module and Hash2 Module are deployed to generate the two indices hashed from physical address.
Each entry can be inserted to any candidate set as long as it has a vacancy.
It should be noticed that \textit{Hash1}, \textit{Hash2} and \textit{fPrint Hash} satisfy the relationship described in Section \ref{Background} to guarantee that a candidate bucket index can deduce the other candidate bucket index with the result of \textit{fPrint Hash}.

\textbf{fPrint Array} consists of two fields:
The 1-bit \textit{}{Valid} flag indicates whether the corresponding entry is meaningful.
The  \textit{f}-bit \textit{}{fPrint} records the hashed value from physical address of each entry during the insertion.

\textbf{Data Array} consists of one saturation counter: 
Ping-Pong is shaped when \textit{Security} reaches the threshold (secThr = 3). The Auto-Cuckoo filter response the Query of PiPoMonitor by Response this value. A Response of \textit{secThr} informs the latter the Access line behaves in Ping-Pong pattern.
\begin{figure*}[t]
	\begin{minipage}{0.48\linewidth}
		\centerline{\includegraphics[width=0.98\textwidth]{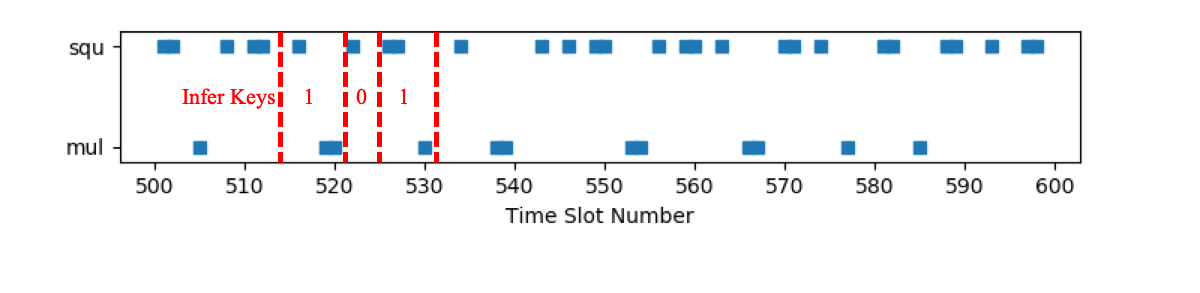}}  
		\vspace{-1.5em}
		\centerline{(a) Baseline}
	\end{minipage}
	\hfill
	\begin{minipage}{0.48\linewidth} \centerline{\includegraphics[width=0.98\textwidth]{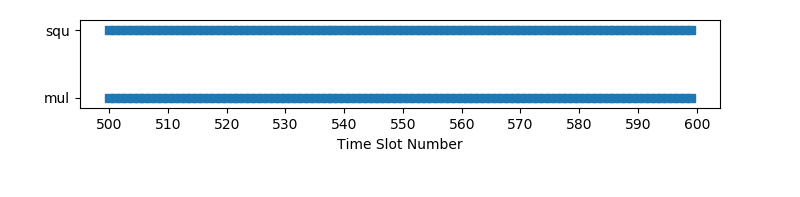}}  
	    \vspace{-1.5em}
		\centerline{(b) PiPoMonitor}
	\end{minipage}
	\caption{Cache usage patterns of probe addresses extracted by the attacker.}
	\label{fig:security}
\vspace{-1.5em}
\end{figure*}

\section{Security Analysis}
\label{SecurityAnalysis}

\subsection{Effectiveness on Ping-Pong Capturing}

We launched Prime+Probe\cite{DBLP:conf/sp/LiuYGHL15} attacks, trying to steal the key from a Square-and-Multiply algorithm (GnuPG version 1.4.13). The algorithm processes the key iteratively from high to low bits, one bit in each iteration. If the bit is 1, \textit{square} and \textit{multiply} performed; otherwise, only \textit{multiply} performed. The sequence of above operations indirectly expose the key.

The attacker and victim run on different physical cores. Every 5000 cycles, the attacker probes two target addresses: the entry addresses of the \textit{square} and \textit{multiple}. In each attack iteration, the attacker accesses eviction sets to evict target addresses from LLC and re-accesses them. When a target address is accessed by victim, an address of the eviction set will be evicted from LLC, causing  attacker's re-access having a large delay due to cache miss; otherwise, all re-accesses hit.

Figure \ref{fig:security} shows the probe results of 100 attack iterations with and without PiPoMonitor. Each blue dot represents that the attacker observed a large delay and inferred the victim may have accessed the target addresses. In Figure \ref{fig:security}(a), the attacker obtained the victim's operation sequence of \textit{square} and \textit{multiply}. When deploying PiPoMonitor, the attacked addresses are recognized as Ping-Pong lines and be protected by Prefetch. In Figure \ref{fig:security}(b), no matter whether the victim has accessed, the attacker always observes accesses, so that the genuine operation sequence cannot be obtained.

\subsection{Defeating Defense-Aware Adversary}

PiPoMonitor relies on the Auto-Cuckoo filter to record memory access traffic. However, due to hardware limitation, the filter cannot record all lines simultaneously. Not surprisingly, during each attack iteration, adversaries can try to deterministically evict the target record from the filter before the victim’s re-accesses shape it into a Ping-Pong.

\textbf{Brute force.} An straightforward manner is to use enormous new addresses to fill the Auto-Cuckoo filter, causing conflicts and evicting the target record. We assume each fill of the adversary can evict one stored record in the filter without the interference of fingerprint collisions. Due to the randomness introduced by autonomous deletion, the probability that the attacker can evict the target record with each fill is:
$$P(evict) = 1/(b\times l)$$ 

Therefore, the mathematical expectation of the fills required to evict a target record is \textit{b * l}. Our experimental results has supported this conclusion, when \textit{b}=8, \textit{l}=1024, we found the adversary needed 8192 memory accesses on average to evict the target record. This severely slowed down the attack, and the time consumed exceeds the iteration interval limit of existing attacks to probe victims' cache accesses.



\begin{figure}[h] 
    \centering
	\includegraphics[width=0.44\textwidth]{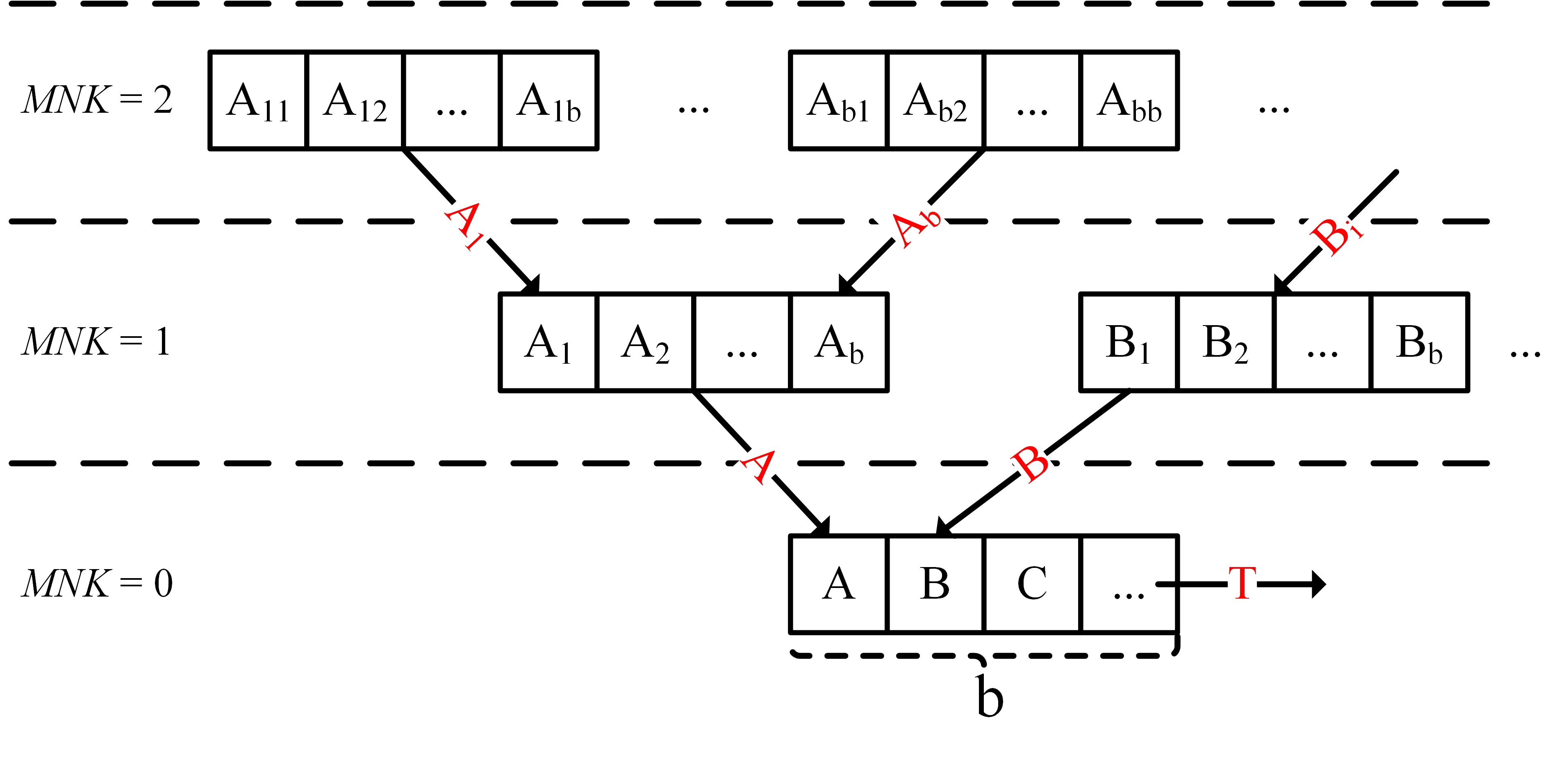}
	\vspace{-1.5em}
	\caption{The reverse attacks on the Auto-Cuckoo filter with different \textit{MNK}}
	\label{fig:reverseAttack}
\vspace{-1.6em}
\end{figure}

\textbf{Reverse engineering attacks.} The more effective method is to construct an eviction set and fill the filter with a small number of addresses. Unfortunately, the adversary have to deal with the \textbf{exponential} impact of \textit{MNK} on eviction uncertainty. Shown by Fig. \ref{fig:reverseAttack}, the target record \textit{T} is located in the bottom bucket with \textit{b} entries. When $MNK=0$, the adversary only needs to fill \textit{b} records $\{A, B...\}$ as an eviction set to evict \textit{T}. Although the filter randomly selects victim to evict in a bucket, the adversary can continue to fill the set and evict \textit{T} in a linear time. When $MNK=1$, only when $\{A, B...\}$ arrives at the bottom bucket after 1 relocation can probably trigger the eviction of \textit{T}. The attacker needs to fill \textit{b} eviction sets $\{\{A_1,A_2...A_b\}, \{B_1,B_2...B_b\}...\}$ to relocate $\{A,B...\}$ from the middle buckets to the bottom to continuously triggers eviction and evicts \textit{T} eventually. By analogy, when $MNK=2$, shown by Fig. \ref{fig:reverseAttack}, the size of the overall eviction set reaches $b^3$. The Auto-Cuckoo filter is configured as $b=8, MNK=4$ in this design, making the eviction set size reaching $b^{(MNK+1)}=32768$. This makes the time cost of the reverse attack even exceed the brute force, rendering it impractical.


\subsection{False Negatives}

Autonomous deletion facilitates the Auto-Cuckoo filter to build a dynamic observation window, but it inevitably introduces false negatives. In fact, the observed address space is extremely larger than the filter itself. It's still competent because the defense goal is to guarantee that within the window, the attacker’s LLC evictions and victim’s re-accesses will shape the target record into Ping-Pong. A reasonable \textit{b $\times$ l} can generally ensure the time required for the record to become Ping-Pong, and sporadic false negatives cannot hinder the eventual capture when the subsequent Access arrives. Most importantly, autonomous deletion makes it impossible to deterministically evict a target record and create false negatives of actual threats.
\section{Evaluations}

\subsection{Methodology}

We implemented PiPoMonitor on Gem5\cite{DBLP:journals/sigarch/BinkertBBRSBHHKSSSSVHW11}, a cycle-accurate simulator. The baseline is configured as a quad-core processor with an inclusive cache architecture. Each processor core contains private L1 and L2 cache. The shared L3 cache is physically distributed as slices. The PiPoMonitor is deployed in the memory controller, and mainly consists of the Auto-Cuckoo filter. The detailed parameters are listed in Table \ref{tab:Setup}.


\begin{table}[htbp]
	\vspace{-1.0em}
	\renewcommand{\arraystretch}{1.3}
	\newcommand{\tabincell}[2]{\begin{tabular}{@{}#1@{}}#2\end{tabular}}
	\caption{System configurations}
	\label{tab:Setup}
	\centering
	\scriptsize
	\begin{tabular}{|c|c|} 
		\hline
		\multicolumn{2}{|c|}{\textbf{Baseline Parameters}}\\
		\hline  
		CPU&4 cores at 2.0 GHz\\
		\hline
		Coherence protocol&MESI\\
		\hline
		\tabincell{c}{L1I/L1D caches}&\tabincell{c}{Inclusive, Private, 64 KB, 4-way, 2 cycles}\\
		\hline 
		\tabincell{c}{L2 caches}&\tabincell{c}{Inclusive, Private, 256 KB/core, 8-way, 18 cycles}\\
		\hline 
		\tabincell{c}{L3 caches}&\tabincell{c}{Inclusive, Shared, 4 MB, 16-way, 35 cycles}\\
		\hline 
		DRAM& 200-cycle latency\\
		\hline 
		\multicolumn{2}{|c|}{\textbf{PiPoMonitor Parameters}}\\
		\hline 
		\tabincell{c}{Auto-Cuckoo filter}&\tabincell{c}{\textit{l}=1024, \textit{b}=8, \textit{f}=12, $\epsilon$=0.004, \textit{secThr}=3, \textit{MNK}=4}\\
		\hline 
	\end{tabular}
\end{table}

As shown in the table \ref{tab:mixes}, we composed 10 workloads mixed by benchmarks from the SPEC CPU2006. Each workload contains 4 benchmarks with the reference input size. The benchmarks run concurrently on the quad-core processor. The baseline system runs the same workloads without PiPoMonitor. For each benchmark, we simulate 1 billion instructions in its core stage and compare the overall execution time.


\begin{table}[htbp]
	\vspace{-1.0em}
	\renewcommand{\arraystretch}{1.3}
	\newcommand{\tabincell}[2]{\begin{tabular}{@{}#1@{}}#2\end{tabular}}
	\caption{Workloads used in our evaluation.}
	\label{tab:mixes}
	\centering
	\scriptsize
	\begin{tabular}{cc} 
		\hline
		MIX&Components\\
		\hline
		mix1&libquantum-mcf-sphinx3-gobmk\\
		mix2&sphinx3-libquantum-bzip2-sjeng\\
		mix3&gobmk-bzip2-hmmer-sjeng\\
		mix4&libquantum-sjeng-calculix-h264ref\\
		mix5&astar-libquantum-mcf-calculix\\
		mix6&astar-mcf-gromacs-h264ref\\
		mix7&gcc-milc-gobmk-calculix\\
		mix8&gcc-mcf-gromacs-astar\\
		mix9&h264ref-astar-sjeng-gcc\\
		mix10&gromacs-gobmk-gcc-hmmer\\
		\hline
	\end{tabular}
	\vspace{-1.5em}
\end{table}

\begin{figure*}[t]
	\begin{minipage}{0.46\linewidth}
		\centerline{\includegraphics[width=0.96\textwidth]{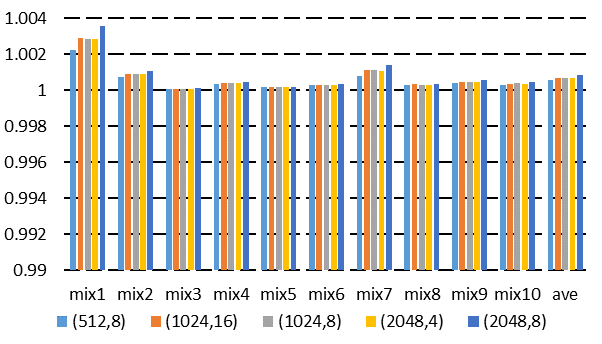}}
		\centerline{(a) Normalized performance}
	\end{minipage}
	\hfill
	\begin{minipage}{0.47\linewidth} \centerline{\includegraphics[width=0.96\textwidth]{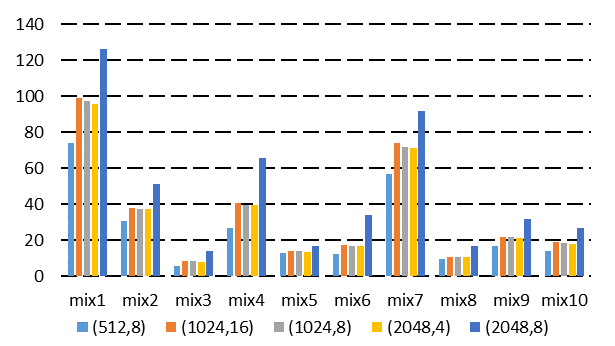}}  
		\centerline{(b) Number of false positives per million instructions}
	\end{minipage}
	\caption{Performance evaluation with different Auto-Cuckoo filter sizes ($l,b$).}
	\label{fig:performance}
	\vspace{-1.5em}
\end{figure*}

\subsection{Performance Evaluation}

Fig. \ref{fig:performance} (a) demonstrates the normalized performance of each workload to the baseline (so higher is better). Taking \textit{l} = 1024, \textit{b} = 8 as an example, the performance is improved by 0.1$\%$ on average. Among them, mix3, mix5 and other test sets have almost no performance changes, and the performance of mix1 has improved the most, reaching 0.3$\%$.

The impact of PiPoMonitor on performance is caused by false positives. In our experiments, all cache lines having a Ping-Pong behavior and triggering Prefetch are considered as false positives. Fig. \ref{fig:performance} (b) shows the number of false positives per million instructions in each mix. 
We can conclude that Prefetch these Ping-Pong cache lines is usually a benefit for performance. For example, when \textit{l}=1024 and \textit{b}=8, there are more false positives in mix1 and mix7, which are respectively 97 and 71 per million instructions. Their performance has also improved the most. On the other hand, the false positives of mix3 and mix6 are less than 20 per million instructions, and their performance is almost unchanged.


\subsection{Sensitivity Analysis}

\textbf{Size of the Auto-Cuckoo filter.} Security analysis has proved that a larger Auto-Cuckoo filter means greater security. At the same time, a redundant filter means that more evicted cache lines can be recorded simultaneously, which may cause concerns about the impact of more false positives on performance. As shown in Figure \ref{fig:performance} (b), we evaluated the Auto-Cuckoo filters configured with different size (512 $\times$ 8, 1024 $\times$ 8, 1024 $\times$ 16, 2048 $\times$ 4, 2048 $\times$ 8) and found that the average impact on performance is less than 0.2$\%$. Therefore, we suggest that the size of the Auto-Cuckoo filter should be considered from the perspective of storage overhead and security.

\textbf{Security Threshold (\textbf{\textit{secThr}}).} The \textbf{\textit{secThr}} affects the sensitivity of PiPoMonitor. When a line’s Ping-Pong traffic exceeds this value, PiPoMonitor considers it suspicious and protects the line. Obviously, the smaller the value, the better it is for identifying abnormal behavior, but it may also generate more false positives. Experimental results affected by different \textit{secThr} have confirmed this: the average performance when the threshold is 3 is better than when it is 1 or 2.

\subsection{Hardware Overhead}

We evaluated the storage and area overhead of PiPoMonitor according to the configuration in table \ref{tab:Setup}. The calculation was performed using CACTI 7 \cite{DBLP:journals/taco/Balasubramonian17} under the 22 nm technology. The Auto-Cuckoo filter contains 1024 $\times$ 8 = 8192 entries. Each filter entry has an \textit{fPrint} field (12 bits), a \textit{Security} counter (2 bits), and a valid (1 bit). Therefore, PiPoMonitor requires additional 15 KB storage overhead compared to the last-level cache, which is 0.37$\%$. It occupies 0.013 $mm^2$, which is 0.32$\%$ more than the LLC. In addition, for a high-performance chip with more cores and larger LLC, the overhead could be further decrease.

\section{Related Work}



Back-invalidation is a common feature for cross-core attacks.
SHARP\cite{DBLP:conf/isca/YanGST17} modifies the cache replacement policy to randomly evict cache-lines while detecting the back-invalidations.
BITP \cite{DBLP:conf/IEEEpact/Panda19} employs a hardware prefetcher to prefetch the back-invalidation lines to interfere the observation of attackers.
RIC \cite{DBLP:conf/dac/KayaalpKEEAPJ17} propose to relax the inclusion for read-only data and thread-private data, which prevent the back-invalidation of these data from other cores.
However, there are large false positive ratios as back-invalidation behaviors happens in normal execution. Wang et al. proposed stateful approaches \cite{DBLP:conf/date/WangYHJM20, DBLP:conf/cf/WangYHLJM19} that extended directories to record cache line coherence events and identified more precise Ping-Pong patterns among cache-memory traffic. However, the storage overhead of the directory extension is essential and the directory itself is vulnerable to reverse attacks using eviction sets to evict target records. In contrast, PiPoMonitor addresses the challenge of a stateful approach by using space-efficient probabilistic hardware, the Auto-Cuckoo filter, to reduce storage overhead and defeat reverse engineering attacks.

\section{Conclusion}
This paper proposes PiPoMonitor, a novel stateful approach to defend against cross-core cache side channel attacks. 
It uses space-efficient probabilistic hardware, the Auto-Cuckoo filter to reduce storage overhead, and enhances layout non-determination against reverse engineering attacks. Evaluations and security analysis show PiPoMonitor achieves the defense goal with minimal performance and hardware overhead.

\section{Acknowledgment}

This work was supported in part by the Strategic Priority Research Program of Chinese Academy of Sciences under grant No.XDC02010000, and in part by the National Scientific Research Project of China under grant No.2019603-001.

\bibliography{refs.bib}
\bibliographystyle{unsrt}

\end{document}